\documentclass[11pt]{article}
\usepackage{acl2015}
\usepackage{times}
\usepackage{float}
\usepackage{url}
\usepackage{latexsym}
\usepackage{graphicx}
\usepackage{amsmath}
\usepackage{chicago}
\usepackage[toc,page]{appendix}
\usepackage{lipsum} 
\usepackage{doi}
\usepackage{hyperref}  
\usepackage{bookmark}  

\title{Precision Design of Cyclic Peptides using AlphaFold}

\author{AU, Cheuk Sau (Jethro)\\
  Hong Kong University of Science and Technology \\ 
  21014200 \\
  {\tt csauac@connect.ust.hk} \\\\}

\date{30/11/2024}

\begin{document}
\maketitle
\begin{abstract}
This independent research investigates methods to improve the precision of cyclic peptide generation targeting the HIV gp120 trimer using AlphaFold. The study explores proximity-based hotspot mapping at the CD4 binding site, centroid distance penalization, generative loss tuning, and custom loss function development. These enhancements produced cyclic peptides that closely resemble the binding conformation of the CD4 attachment inhibitor BMS-818251. The proposed methodology demonstrates improved structural control and precision in cyclic peptide generation, advancing the applicability of AlphaFold in structure-based drug discovery.
\end{abstract}

\section{Introduction}
Recent advances in artificial intelligence have catalyzed major breakthroughs in structural biology, particularly in protein engineering and drug discovery. AlphaFold2, developed by DeepMind, achieved unprecedented accuracy in the 14th Critical Assessment of Structure Prediction (CASP14) (Jumper et al., 2021). Despite its success, adaptations of AlphaFold for cyclic peptide design targeting the HIV gp120 trimer—especially its CD4 binding site (CD4bs)—remain underexplored. This study seeks to address this gap by enhancing AlphaFold’s generative precision for cyclic peptide design.

During the summer term project, an independent study was conducted to configure the AlphaFold network to generate cyclic peptides for various protein structures. In this previous study, the network successfully generated cyclic peptides targeting short-protein sequences with low RMSD values and comparable to literature generations. However, when the same network was applied to the larger gp120 trimer, the previous model exhibited high variability in the binding region, produced low-confidence or infeasible structures, and was largely unoptimized in the generation process (Figure 1). In response, a further research project was initiated to improve the precision of AlphaFold’s generative capabilities and its ability to target the HIV gp120 trimer. 
\begin{figure}[H]
    \centering
    \includegraphics[width=1\linewidth]{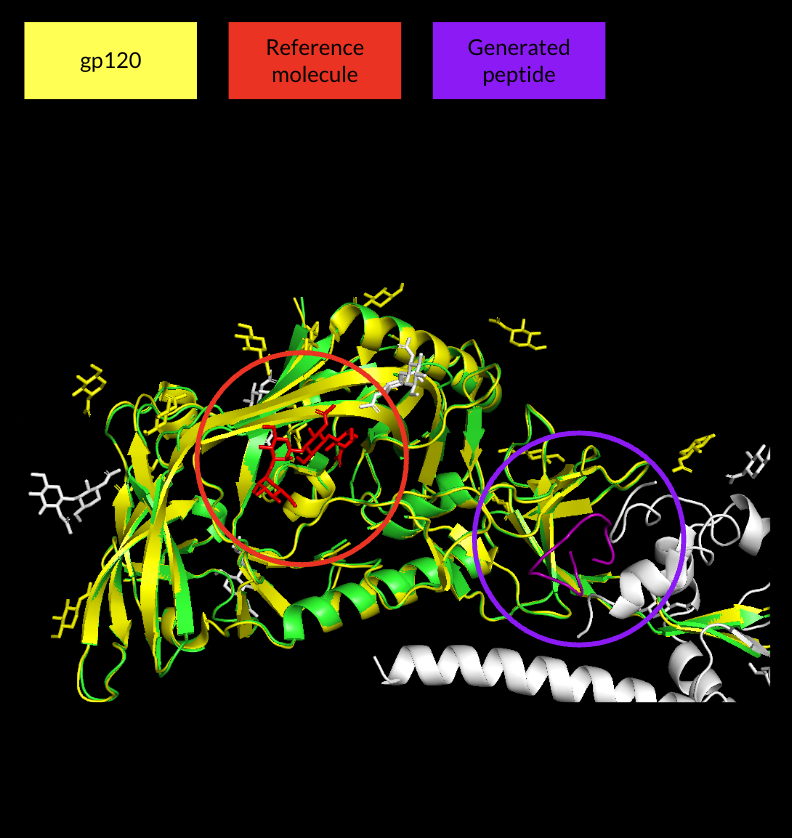}
    \caption{Previous peptides generated were not precise (generated in purple does not conform close to the reference molecule in red)}
    \label{fig:1}
\end{figure}

\section{Related Work}
This study builds upon work on adapting the AlphaFold2 model to generate cyclic peptides (Rettie et al. and Kosugi et. al). In both of their work, a circular distance embedding was encoded into the input embedding of the AlphaFold network, successfully tuning the AFDesign implementation of AlphaFold2 (Mirdita et al.) to generate cyclical peptides with high accuracy. To compare the accuracy of AlphaFold2’s targeting capabilities, the BMS-818251 molecule and its crystal structure were used as references (Lai et. al). Furthermore, DeepMind has recently released its latest iteration of AlphaFold, AlphaFold3, which has extended the generative capabilities of AlphaFold to beyond protein structure to ligands, DNA, RNA, and more (Abramson et.al). However, as this iteration of the model was not open-sourced at the time this research was commenced, only the AlphaFold2 open-sourced was used. 

\section{Exploring peptide control with Hotspots}
AFDesign offers more nuanced control during the peptide generation process through a hotspot configuration. This configuration allows the user to provide a string of residue numbers the model’s loss generation focuses on. During the generation process, the hotspot configuration masks out non-interest residues when calculating the interface contact scores, causing the model to only consider interfaces near the hotspot residues.
\subsection{Improving control over cyclic peptide generations}
To first explore the impact of hotspot configurations on the final conformation of cyclic peptides, a series of cyclic peptides with \textit{binder length=13,} \textit{seeds=} \textit{[1, 2, 3],} and varying hotspots of \textit{[None, “100-105”, “150-155”, and “200-205”]} were generated using the HIV gp120 trimer (PDB: 6V0R) as template.  
\begin{figure}
    \centering
    \includegraphics[width=1\linewidth]{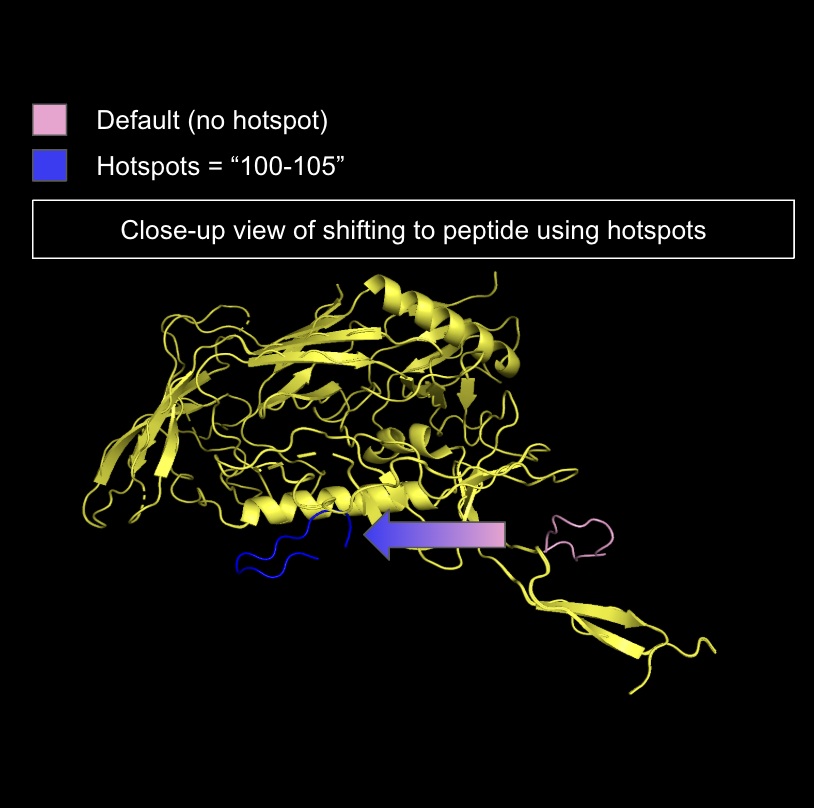}
    \caption{Hotspot configuration experiment}
    \label{fig:2}
\end{figure}
Based on Figure 2 and 3, it can be seen that varying the hotspot configuration can change the final conformation and location of the cyclic peptide. The location of the final cyclic peptide is consistent across seeds for residues \textit{“100-105”} and \textit{“150-155”}, but the residue region of \textit{“200-205” }varied quite drastically (as seen in Figure 4 in orange). Upon closer examination, it can be seen that the residue corresponding to the \textit{“200”} hotspot configuration is contained in an inner region of the gp120 trimer, and the existing generation method is unable to determine an optimal configuration for a cyclic peptide close to the interfacing region. Upon inspection, we noticed that using the default weights of AFDesign, the generated cyclic peptide tends to remain in the outer domains rather than identify a fitting sequence inside of the trimer.  However, as the residues were arbitrarily determined in this experiment, it also cannot be determined if the selected residue just happens to be an unboundable region, causing the residue to bind to a minima near the region. 
\begin{figure}[H]
    \centering
    \includegraphics[width=1\linewidth]{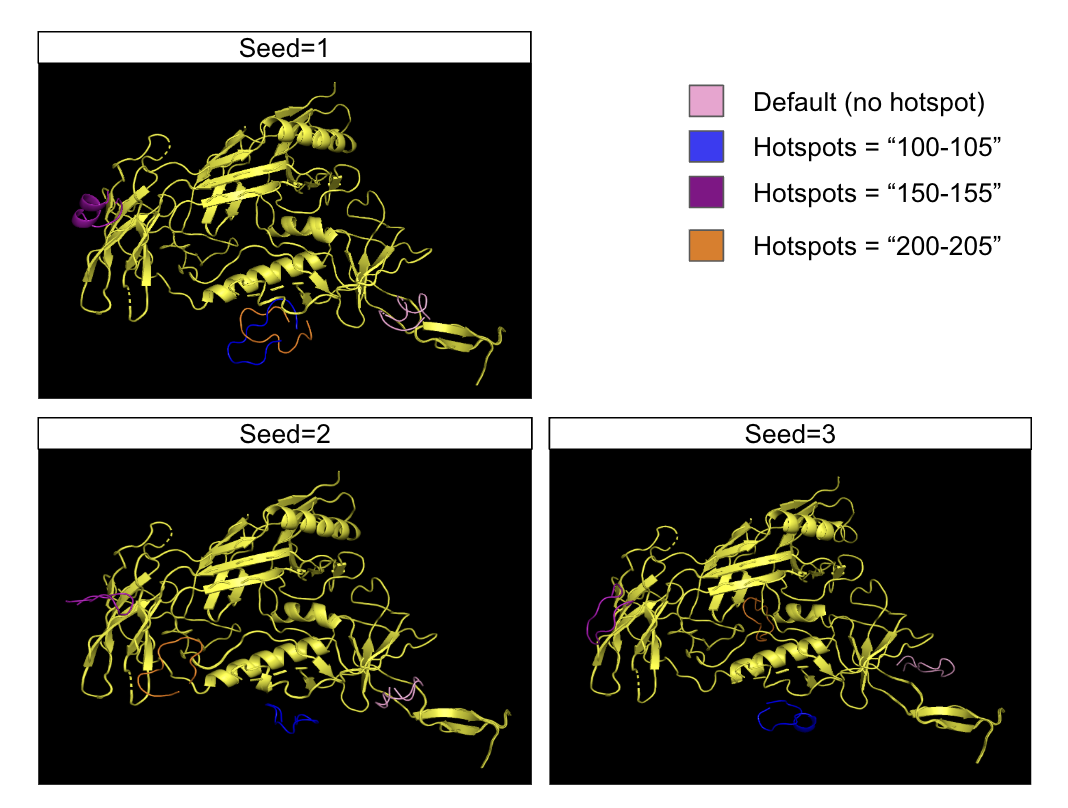}
    \caption{Cyclic peptides generated by varying hotspot location}
    \label{fig:3}
\end{figure}
\begin{figure}[H]
    \centering
    \includegraphics[width=1\linewidth]{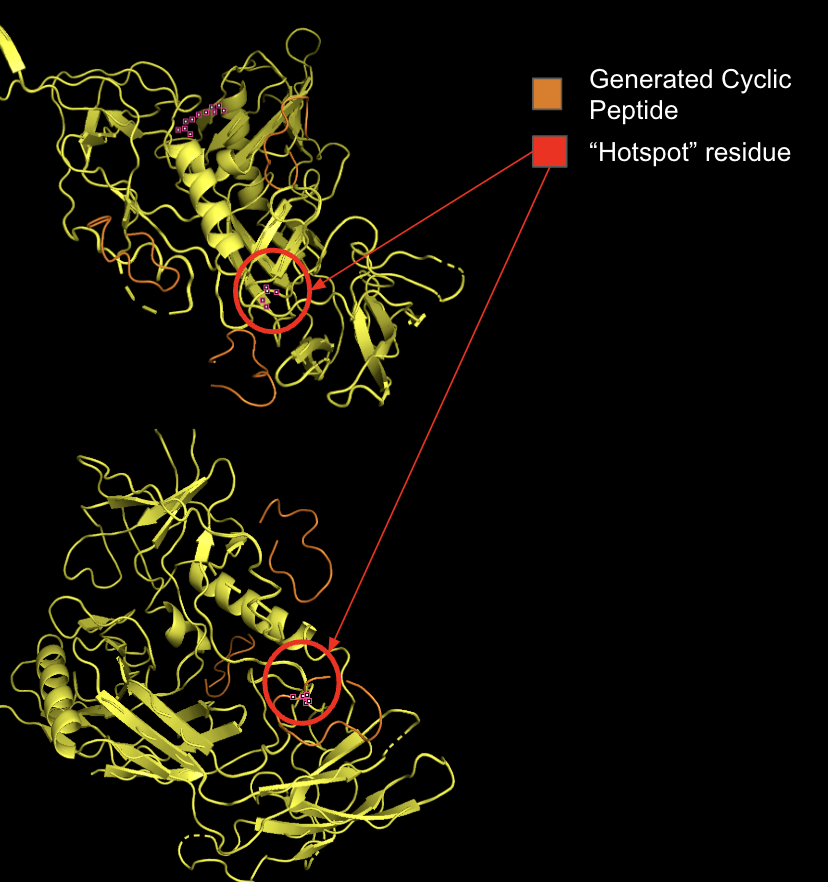}
    \caption{A hotspot residue within the inner region of HIV caused variations in resulting peptide binding location}
    \label{fig:4}
\end{figure}

\subsection{Proximity-based configuration of hotspots for gp120 CD4-binding site (CD4bs)
}
Further exploration was conducted to better understand the peptides generated through hotspot configurations. In the previous experiment, the hotspot residues were arbitrarily selected. On the contrary, in this experiment, an experimental structure of the attachment-inhibitor molecule BMS-818251 was chosen as a reference conformation (PDB: 6MU7). The PDB 6MU7 file includes a crystal structure of the gp120 trimer bound to a small molecule at the CD4-binding site, effectively serving as an antiretroviral drug. Using the BMS-818251, a known attachment inhibitor that binds to the CD4bs, we can use the molecule as a proxy to map and identify the target hotspot residues. 

A proximity-based method was used to identify the hotspots corresponding to the reference molecule. In this method, residues within proximity of 5-Angstrom from the reference BMS-818251 molecule were mapped as hotspots (the molecule was selected and filtered using the ‘JYY’ chain-id). Then, using the proximity-mapped hotspots, additional cyclic peptides were generated. From the generated peptides, the conformation closest to the reference molecule was selected for inspection. The proximity hotspot mapping and resulting generations are in Figure 5 below. 
\begin{figure}[H]
    \centering
    \includegraphics[width=1\linewidth]{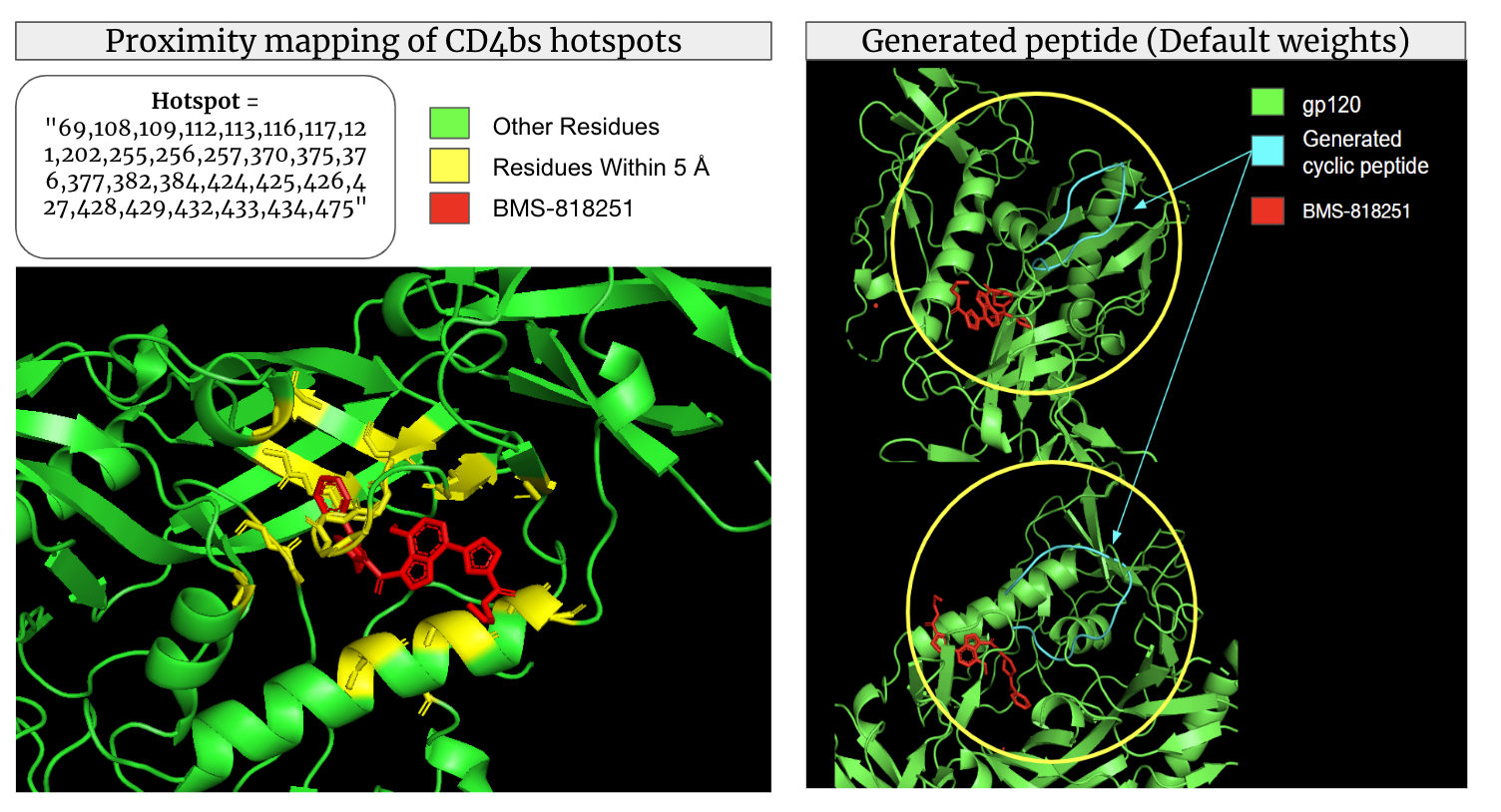}
    \caption{Proximity mapping in hotspot increases precision of generated peptide}
    \label{fig:5}
\end{figure}

Referring to Figure 5, the generated cyclic peptide was \textit{close} to the CD4bs but not within the CD4 pocket, where the attachment inhibitor is. Similar to the case before, as the CD4bs includes an inner region of the gp120 trimer, the resulting locations of the cyclic peptides varied quite drastically, similar to that from our initial hotspot generation we noticed above. Given that the AlphaFold2 model is trained to minimize the energy of protein conformations, it is hypothesized that during the generation process, the model is unable to overcome the high energy barrier posed by the outer domain of the trimer. This implies that during the generation process, the model opted to accept a reduction in CD4 contact over overcoming this higher energy barrier, resulting in a ‘lazy’ structure. To further improve the peptide generation accuracy, there needs to be a greater penalization of distance, leading to research into modifying AFDesign’s loss function.

\section{Improving peptide generations in the CD4 region through loss function}

Recall the AlphaFold cyclic peptide generation process involves a loss function based on the AlphaFold models intermediary outputs: \textit{pLDDT}, \textit{PAE}, and \textit{C-alpha distogram} (Kosugi et. al). In the past generations from the summer term, the original weights in the loss functions were used, namely: \textit{i\_con = 1, pLDDT = 0.1}. Whilst this weighting was sufficient to recreate cyclic peptides from literature in the first phase, it was insufficient to generate cyclic peptides for HIV gp120, as the peptides often were not generated within the target regions. Therefore, to improve on the effectiveness of the generation process, the AlphaFold loss function was modified to penalize generations far away from the CD4bs. 

\subsection{Distance-penalization through centroid-MSE loss}

Since the AFDesign model outputs the XYZ coordinates of the cyclic peptide as part of its intermediary outputs (‘aux’) during each forward pass, the AFDesign loss function could be modified to incorporate an MSE loss (Figure 6) that serves as a regularization parameter in the backpropagation process. By doing so, we could leverage AlphaFold2’s prediction capabilities to calculate a gradient update that alters the peptide conformation closer to the hotspot region (in this case the CD4bs).

\begin{figure}[H]
    \centering
    \includegraphics[width=1\linewidth]{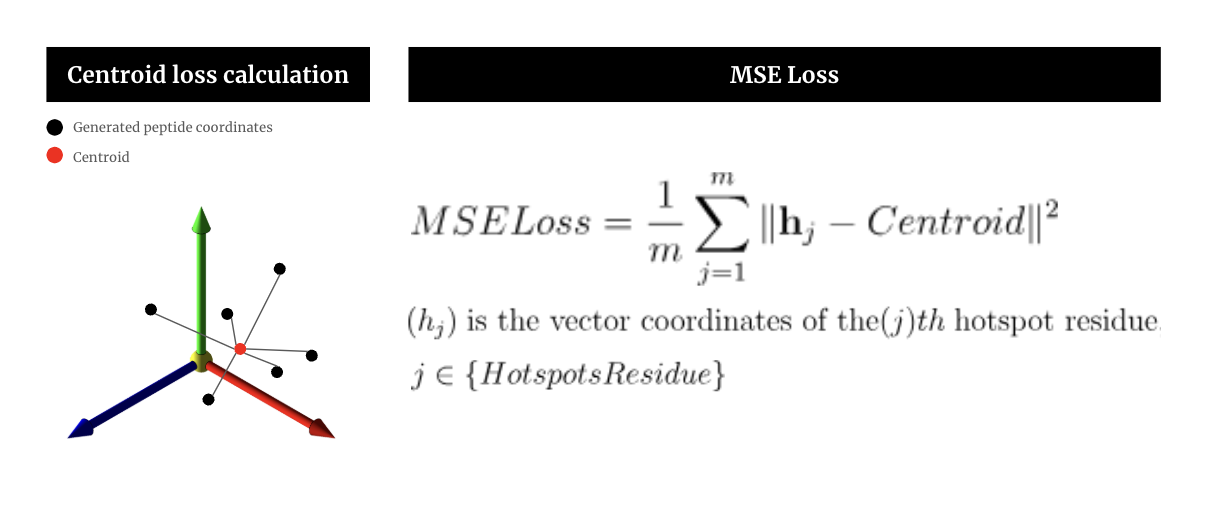}
    \caption{Centroid-distance penalization of generated peptides}
    \label{fig:6}
\end{figure}

To calculate the MSE loss, the generated cyclic peptide is first assumed to be a point cloud of 3-dimensional vectors. Only the point cloud's centroid is used in the loss calculation to maintain the cyclic peptide's protein folding capabilities. The centroid of the generated peptides is calculated during each forward pass by taking the mean of the cartesian coordinates of each of the peptide’s residue positions. The MSE loss is then calculated between the cyclic peptide centroid and all hotspot residues in the configuration. The rationale for configuring a loss in such a manner is to provide the model with flexibility for the torsion angles to rotate and conform to the CD4bs. The MSE loss functions are implemented through JAX wrapper. The MSE loss was set with weight=1.0 and incorporated into the generation process. Additional peptides were generated using seeds from 51 to 54, inclusive, and binder length=6. The results are shown in Figure 7 below. 

\begin{figure}[H]
    \centering
    \includegraphics[width=1\linewidth]{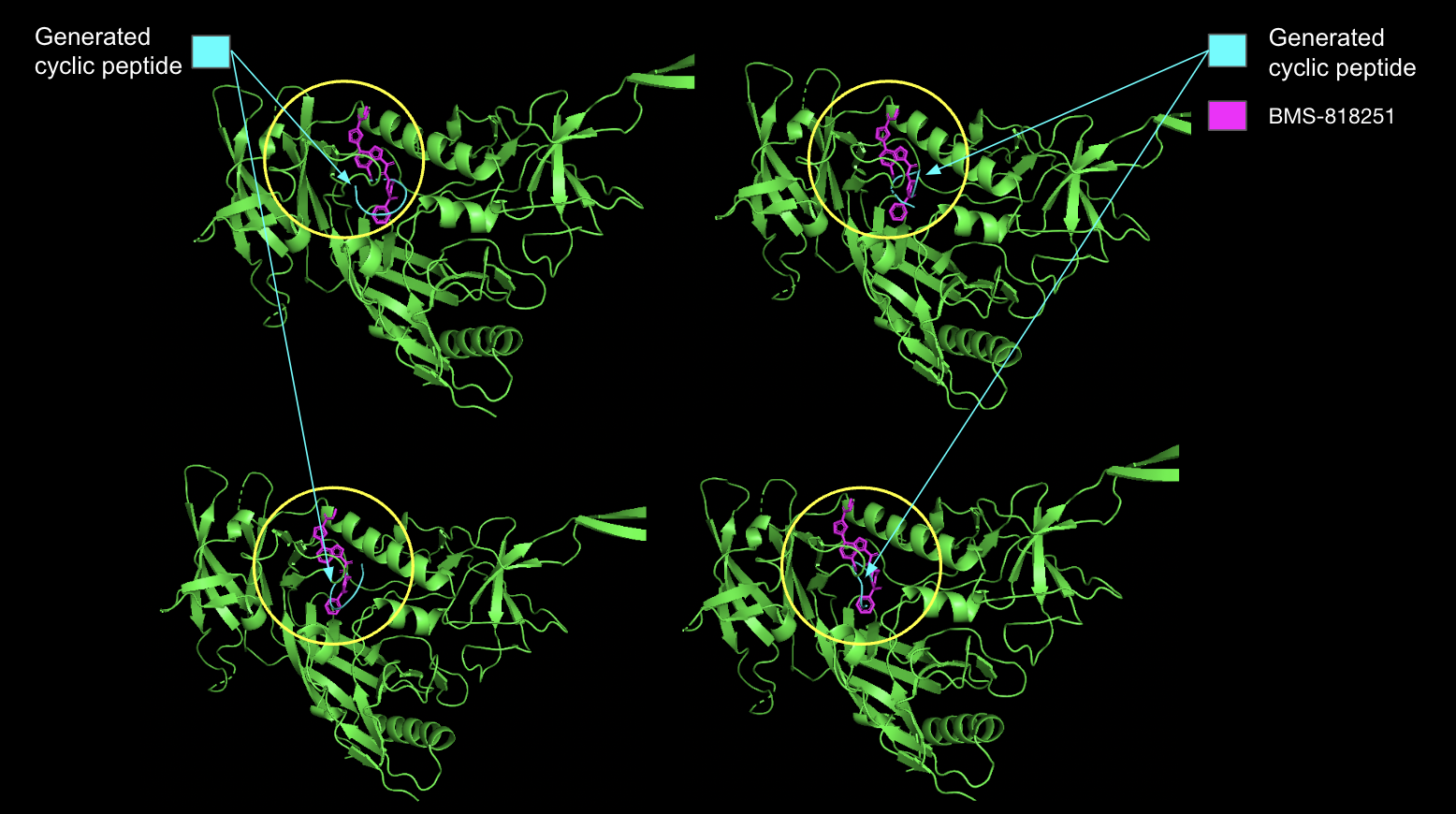}
    \caption{Applying a centroid distance penalization allows the generated peptide to conform directly at the CD4bs}
    \label{fig:7}
\end{figure}
Referring to the figure, generating an additional MSE loss caused the AFDesign model to generate cyclic peptides at the CD4 region. This suggested that the MSE loss operated as intended. However, as the MSE loss was large, it caused overfitting to the CD4bs, and the residue confidence scores (pLDDT) were low as a result. 

\subsection{Improving confidence and binding energy of cyclic peptides }

An exploration of optimal generation to improve the confidence score was then conducted to optimize generation parameters for gp120. A series of additional loss functions were implemented to improve the confidence scores.

\subsubsection{Developing custom loss through \textit{pDockQ} }

Protein-protein interactions are often modeled through docking algorithms to produce scores that help screen high-potential compounds for further evaluation. Recent studies by Pozzati et. all have shown that the use of the commonly-used \textit{DockQ} score can improve the prediction of protein-protein interactions of AlphaFold. However, Pozzati implemented the \textit{DockQ} mainly to evaluate proteins of known sequences; there has been no implementation done when applied to cyclic-peptide generation (Figure 8). 
\begin{figure}[H]
    \centering
    \includegraphics[width=1\linewidth]{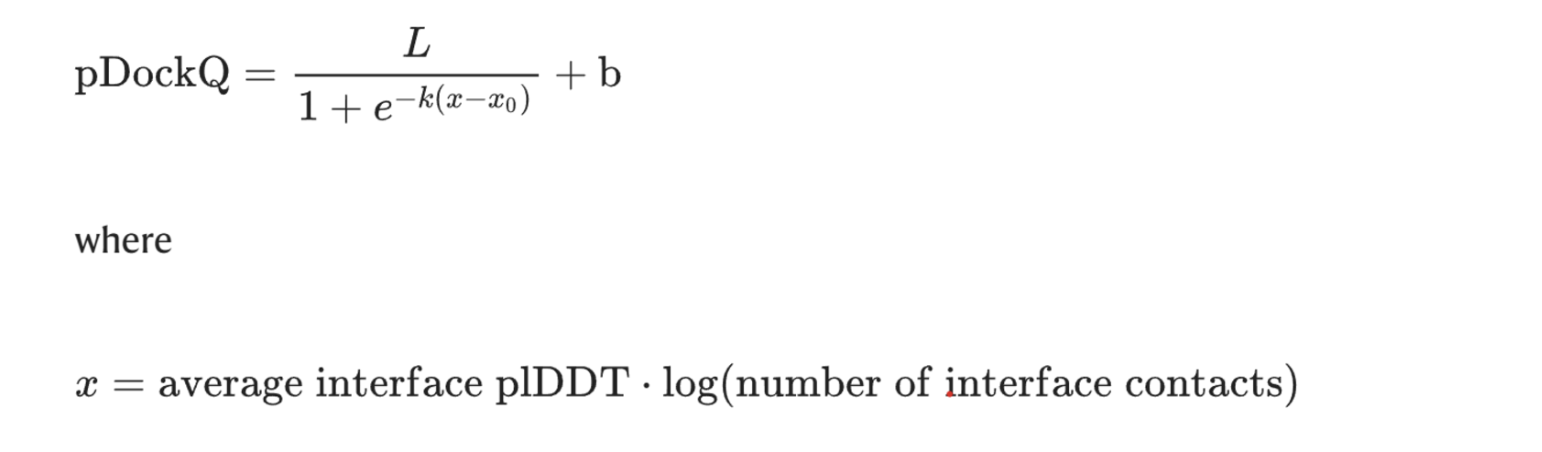}
    \caption{Predicted DockQ scoring from Pozzati et al. }
    \label{fig:8}
\end{figure}
One main challenge with implementing the \textit{DockQ} score is that the intermediary outputs of the AlphaFold2 model are insufficient to calculate the entire score. However, in the study by Pozzati et al., they were able to predict the \textit{DockQ} score through a \textit{pDockQ} score through intermediary outputs from AlphaFold and demonstrate its effectiveness in predicting acceptable and high-quality predictions. The \textit{pDockQ} score is a sigmoid function of the product of the average pLDDT, log number of interface contacts, and a bias term \textit{b}.

Inspired by this, an implementation of the \textit{pDockQ} function was carried out, customizing the function for cyclic peptide generation purposes. A custom loss function of the \textit{pDockQ} score was wrapped in JAX to be incorporated into the backpropagation process during generation for evaluation. During every forward pass, the average \textit{pLDDT} of the binder residues is calculated, and the log number of interface contacts is calculated from the interface contact map variable \textit{i\_cmap}, which outputs a sigmoid of the binned distances between the residues. To determine the interface contact a threshold of: \[\textit{i\_cmap} \geq 0.95\] Figure 9 shows an example of the interface contact map used to calculate the interface contacts between a cyclic peptide of length 13 and the gp120 trimer. 

\begin{figure}[H]
    \centering
    \includegraphics[width=1\linewidth]{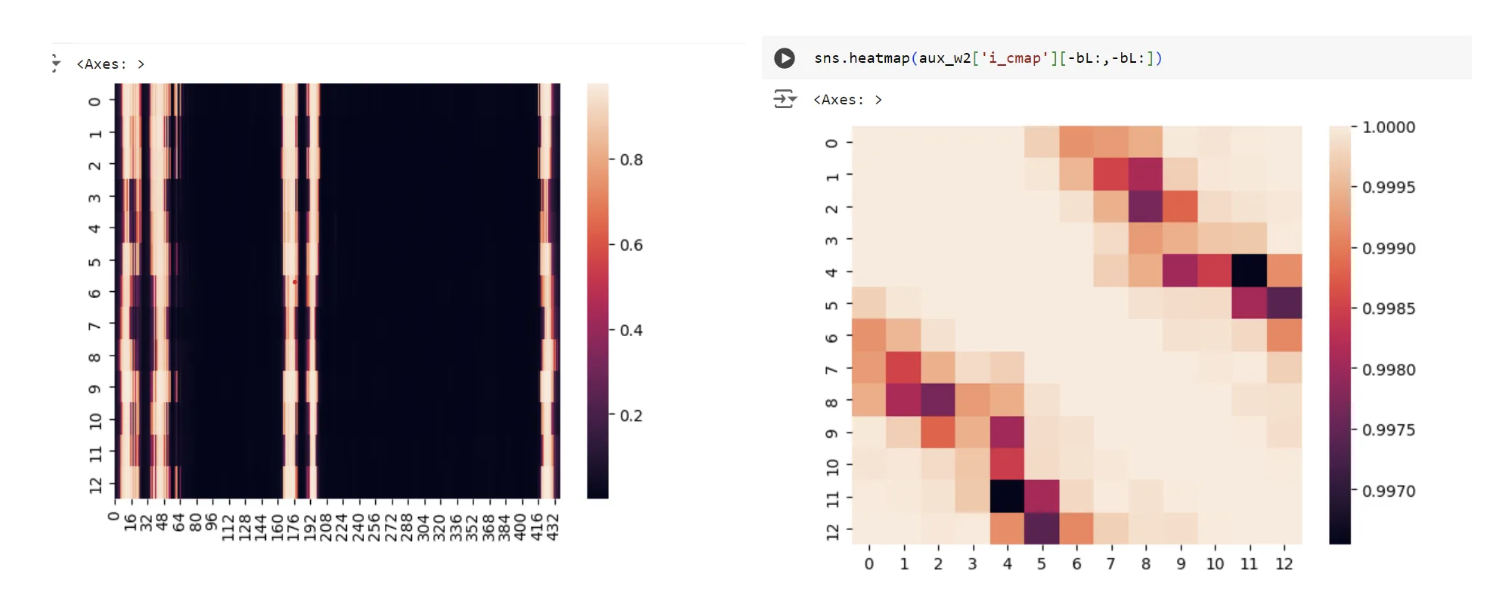}
    \caption{i\_cmap of cyclic peptide of length 13 to gp120 trimer}
    \label{fig:9}
\end{figure}

\subsubsection{Increasing pLDDT weights in the loss function}

Modifications to the \textit{pLDDT} loss function were also considered during this exercise. This is inspired by Rettie et al.’s use of the \textit{pLDDT} as selection criterion in their work to evaluate peptides—models with loss functions of \textit{pLDDT} weightings 0.5 and 1 corresponding to models \textit{w1} and \textit{w2,} respectively.  

\subsubsection{Comparing generative performance across model loss variations}

To compare performance, cyclic peptides of length 13 were generated for \textit{seeds = [20,23,25,26,27]} for 7 models. Generation results for each of the 5 seeds can be seen at Appendix 2, and the selection of the seeds was arbitrary. For each model, the generative loss was computed. As the starting seed influences the scores, the scores for each seed were normalized by the baseline results to evaluate the effectiveness of each hyperparameter setting (Appendix 3). As we are looking to independently assess the effectiveness of these losses, the hotspot MSE loss was not used in this experiment. The details of each model can be seen in Figure 11 below. 

\begin{figure}[H]
    \centering
    \includegraphics[width=1\linewidth]{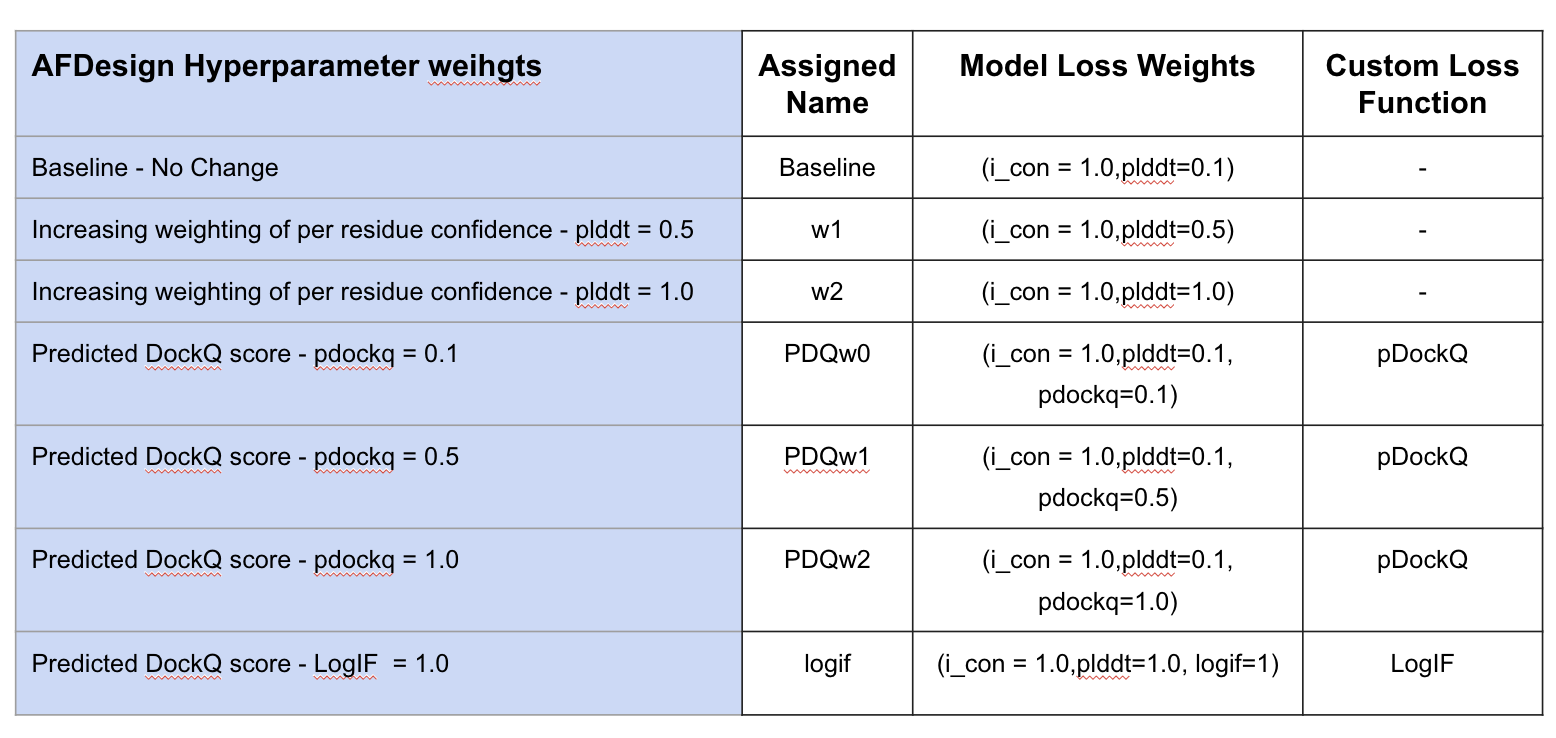}
    \caption{Overview of variations of model losses used }
    \label{fig:11}
\end{figure}

During the evaluations of the \textit{pDockQ}, it was discovered that the custom-loss converges quickly to the maximum value of the \textit{pDockQ} score. The sigmoid function flattens out at the top, leading to a zero gradient during subsequent backpropagation cycles. Given that a higher \textit{pDockQ} score correlates to a better conforming model and the sigmoid function is a monotonically increasing function, a second implementation was evaluated by removing the sigmoid function and retaining the product of the average pLDDT and the log interface contacts score. This loss function is named logIF function, and the model's performance was appended to the evaluation charts (with \textit{logIF} weight = 1.0).

Referring to the normalized generative loss results, it can be seen that increasing the \textit{pLDDT} score weighting and using the \textit{LogIF} function consistently increase the performance of cyclic peptides relative to the baseline (red dashed line). As expected, increasing the \textit{pLDDT} weighting most notably increases the \textit{pLDDT} increase (\textit{pLDDT\_delta}). Interestingly, increasing the \textit{pLDDT} weights also consistently improved the\textit{ i\_ptm} (higher better), \textit{pae} (lower better), and\textit{ i\_con} (lower better) scores. This suggests that increasing the weighting per residue confidence positively interacts with other variables during the generation process. On the other hand, \textit{pDockQ} scores and their variations performed worse, with only marginal improvements relative to the baseline. 

To evaluate the energy levels and solvent-accessible surface areas, the binding energy normalized by SASA score (dG\_separated/SASA * 100) was then computed for all generations using PyRosetta. The generations were minimized using the InterfaceAnalyzerMover package, and peptides were scored using the REF2015 scoring functions [Alford et al.].

Increasing the \textit{pLDDT} weighting reduces the\textit{ dG\_separated/SASA} score (making it more negative), suggesting a more favorable binding. Increasing the \textit{pLDDT} weight to 1 seems to reduce the variation of scores quite drastically compared to a \textit{pLDDT} weight of 0.1. For the \textit{logIF} function, although the variation in the binding energy decreases, the average binding energy does not particularly decrease. Lastly, using \textit{pDockQ} scores caused the model to vary in binding energy. Considering both the generative losses and the PyRosetta energy scores, it can be seen that increasing the \textit{pLDDT} weighting most consistently increases the binding and confidence scores of cyclic peptides. 
\begin{figure}[H]
    \centering
    \includegraphics[width=1\linewidth]{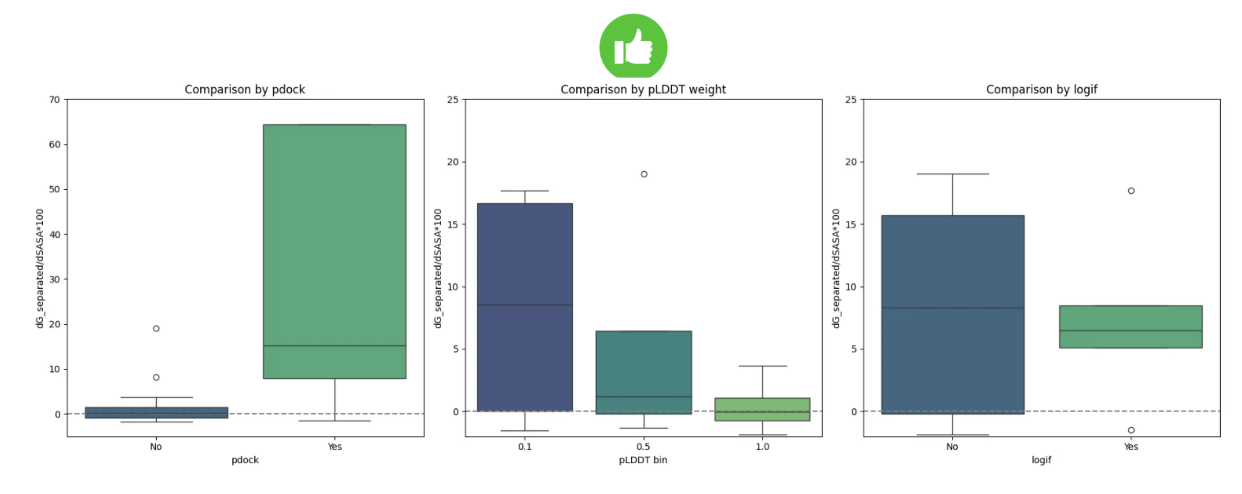}
    \caption{Binding energy normalized by SASA comparison charts}
    \label{fig:12-1}
\end{figure}
\begin{figure}[H]
    \centering
    \includegraphics[width=1\linewidth]{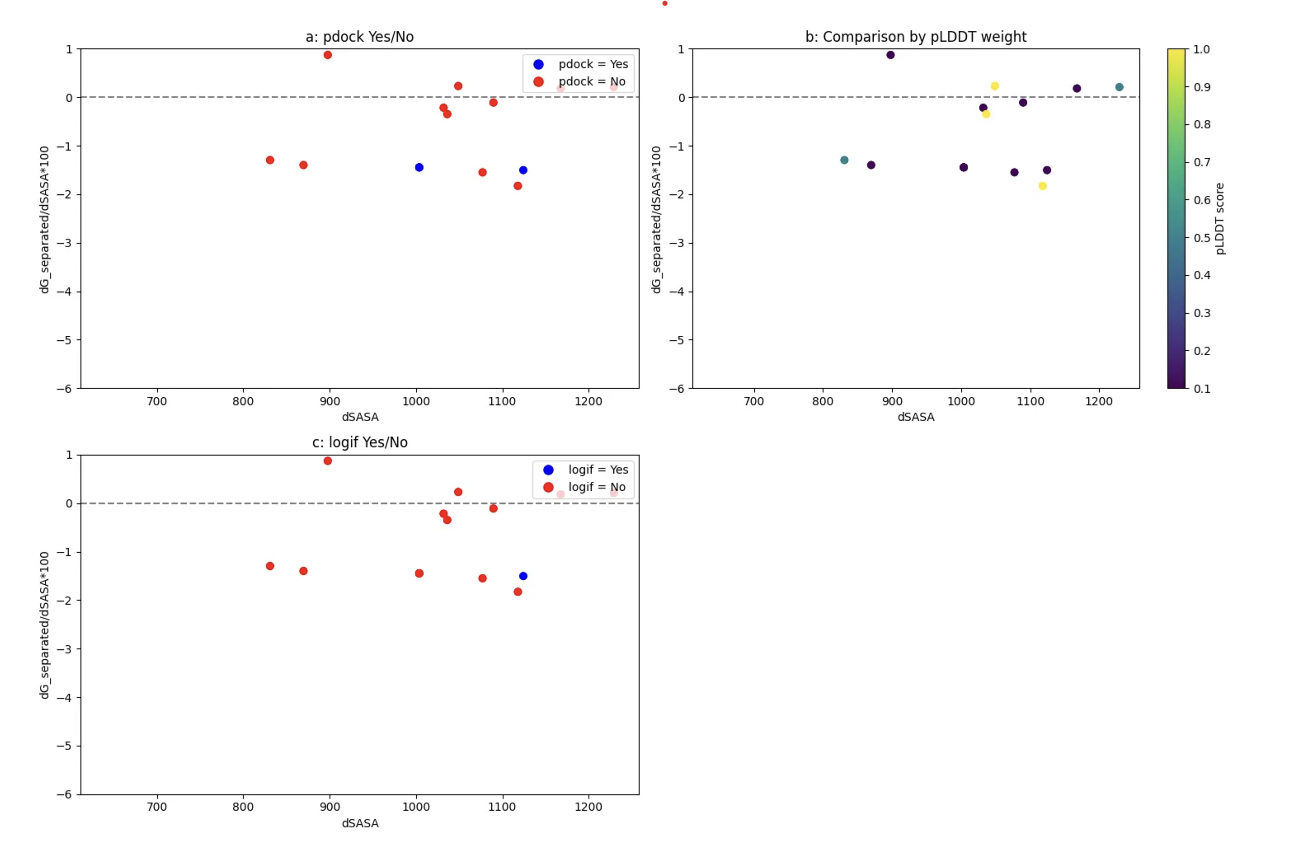}
    \caption{Binding energy vs. SASA plots}
    \label{fig:12-2}
\end{figure}

\section{Precision generation of peptides for gp120 CD4 binding site}

Leveraging all the insights from the above experiments, a final model was developed to generate high-confidence cyclic peptides to target the CD4bs of the HIV gp120 trimer through a series of enhancements. Along with the model's cyclic embedding and hotspot proximity mapping, as mentioned in the above sections, two novel variations were implemented to the model: a confidence-weighted centroid distance penalty and a stepwise weight variation. 

\subsection{Confidence-weighted centroid distance-penalization loss }

Due to the MSE loss causing the model to overfit, the MSE score was normalized by the \textit{pLDDT} score. The rationale was to boost the gradient update of the MSE loss when the \textit{pLDDT} confidence is high while decreasing its significance when the pLDDT is low. Two variations of the \textit{pLDDT} weighting were performed, as seen below in Figure 13. 

\begin{figure}[H]
    \centering
    \includegraphics[width=1\linewidth]{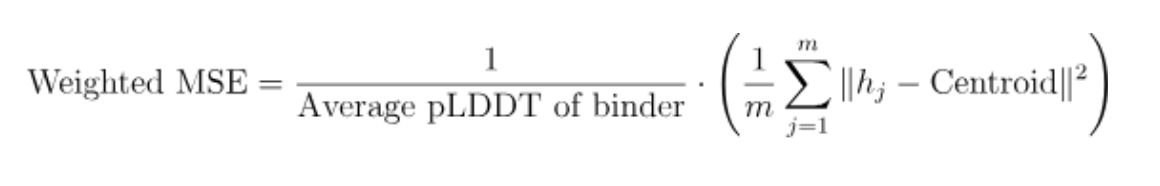}
    \caption{Weighted MSE Loss}
    \label{fig:13}
\end{figure}

In the first implementation, the average \textit{pLDDT} of the binder was calculated and used to normalize the MSE score. However, during a few trial runs, it was noticed that while this reduced the overweighting of MSE, the confidence score was still much less. As a result, a second implementation was conducted in which the \textit{pLDDT} weighting was boosted by a factor of 2, as seen below. This was because as the model switched to one-hot encodings in the later stages of generation, the confidence scores varied quite drastically, and often, a single change to a residue sequence would cause the model confidence to spike upwards. An example of this uptick that caused the cyclic peptide to snap in place with higher confidence can be seen in Appendix 4. To incentivize the model to capitalize on this nonlinear increase in confidence scores, a second loss function was developed using a 2nd-order \textit{plddt} factor. 

\begin{figure}[H]
    \centering
    \includegraphics[width=1\linewidth]{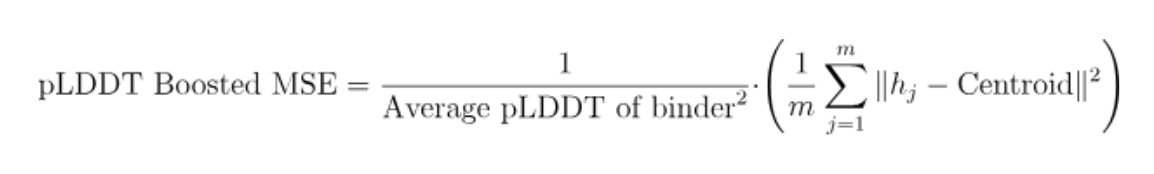}
    \caption{pLDDT boosted MSE Loss}
    \label{fig:14}
\end{figure}

\subsection{Step-wised weight variation}

During the generation process, AFDesign offers the functionality to change loss weightings. Given the challenges in previous sections, a custom design function was designed to vary the weights throughout the 3-stage generation process dynamically. To first have the peptide hone-in on the CD4bs, a weight of \textit{i\_con = 1.0, plddt = 1.0, mse\_plddt\_boosted = 2.0, }was used for the first 60 generation soft-weight generation steps. Afterward, the plddt weighting was increased to a higher value from \textit{1.0 to 5.0 }for the remaining 40 soft and 32 hard generations to increase the confidence within the binding pocket. From trial runs it can be seen that varying the weights throughout the generations changes the resulting conformation and binding location of the peptide. In this exploration, only a simple step-wise change was implemented, but further exploration into how these weights can be dynamically tuned during the generation with feedback can be further explored in future studies. 

Using the above combination, 168 cyclic peptides targeting the CD4bs were generated for binders of lengths 5, 6, 7, and 8. Models with less than 0.45 pLDDT score were filtered, and the remaining were visually inspected to see if they conformed to the CD4bs. For binder lengths $\geq 8$, mean pLDDT values dropped below 0.25, indicating low structural confidence and suggesting that longer cyclic peptides may not fit within the CD4 binding pocket. The remaining cyclic peptides with side chains with the highest overlap with BMS-818251 were shortlisted and shown in Figures 15, 16, and 17 below. 

As seen from the generations, there is a high overlap region between the generated cyclic peptide (in cyan) and reference molecule (in red). This demonstrates that the model parameters successfully achieved precision generation of cyclic peptides at the CD4bs. 
\begin{figure}[H]
    \centering
    \includegraphics[width=1\linewidth]{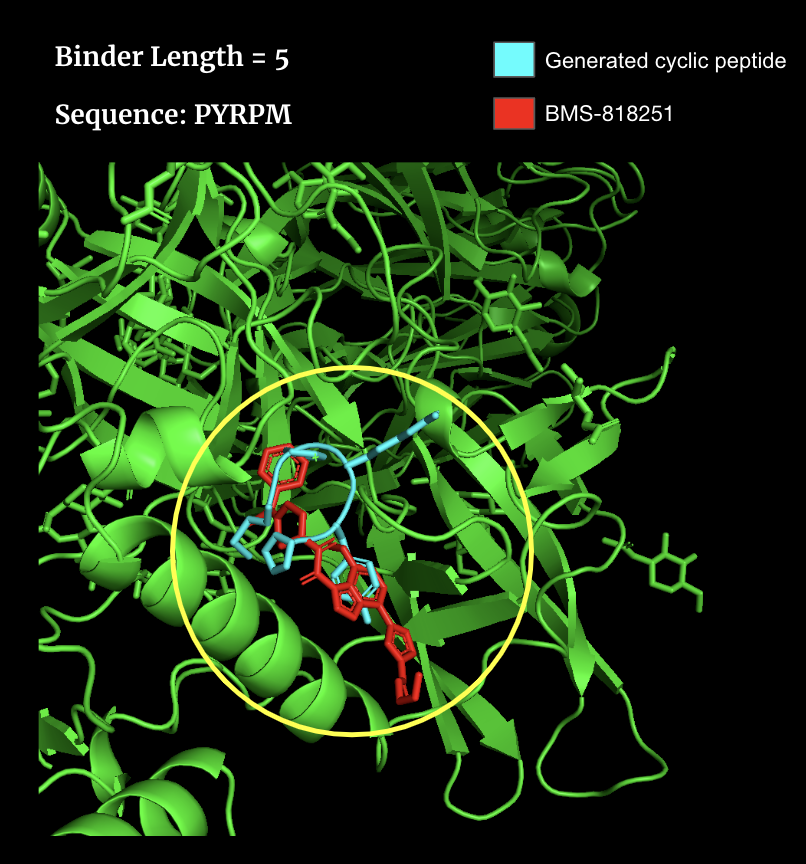}
    \caption{Precision generation of CD4bs cyclic peptides of binder length=5}
    \label{fig:15}
\end{figure}
\begin{figure}[H]
    \centering
    \includegraphics[width=1\linewidth]{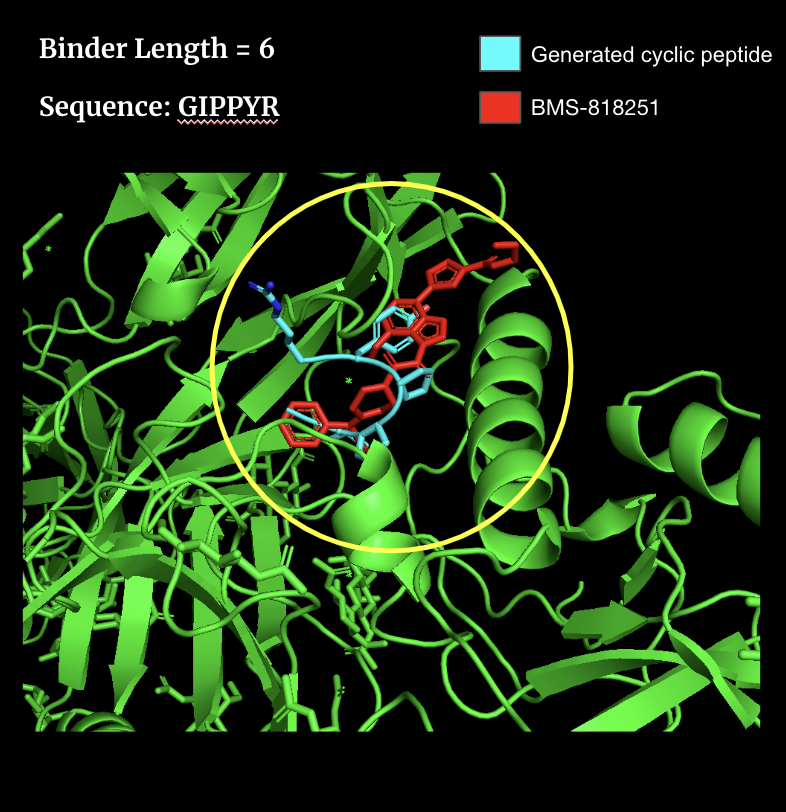}
    \caption{Precision generation of CD4bs cyclic peptides of binder length=6}
    \label{fig:16}
\end{figure}
\begin{figure}[H]
    \centering
    \includegraphics[width=1\linewidth]{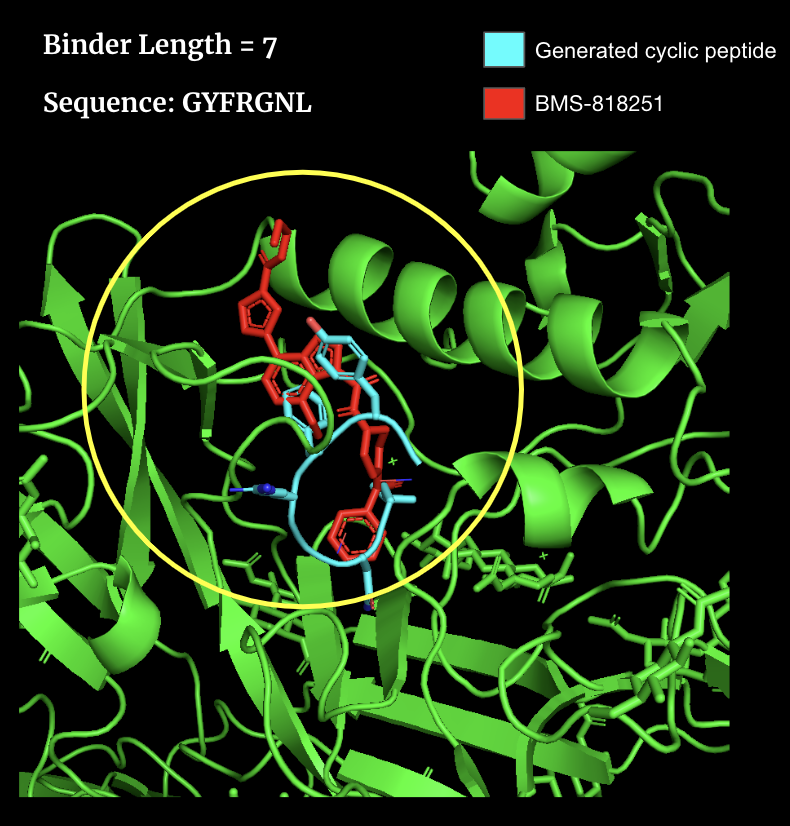}
    \caption{Precision generation of CD4bs cyclic peptides of binder length=7}
    \label{fig:17}
\end{figure}

\section{Limitations \& Future Work}
There are several areas where this study requires further investigation. This study currently only uses   the AlphaFold2 implementation and not the latest AlphaFold3 model, which was recently open-sourced in November 2024.  While the methods explored here can be applied to the AlphaFold3 model, the findings of this study may not directly translate to improvements with AlphaFold3. Moreover, AlphaFold3's enhanced generative capabilities could significantly influence the resulting cyclic peptide generation due to the variations in precision that come from switching foundational models. 

Inspired by the work from HighFold (Zhang et al.), further explorations can be conducted to incorporate disulfide bridges into the generation process. This could improve the stability of cyclical peptides and also offer a more precise approach on where specific disulfide bridges could be attached. Finally, further exploration of model optimization through quantization  could be explored to optimize and accelerate AlphaFold's generations without heavily-compromising accuracy.

\section{Conclusion:}

In this independent study, a series of explorations were conducted to assess methods to improve the precision of generated cyclic peptides targetting the CD4 binding site of the HIV gp120 trimer. Firstly,  a proximity mapping approach was used to identify the hotspot configuration of the CD4 binding site in AlphaFold. Then, a centroid-loss penalization was introduced to enhance the peptide's precision to conform within the CD4bs.  To address the issue of the penalization loss causing overfitting, a further exploration was conducted to evaluate how variations in a model's generative loss could impact the confidence and binding levels of the resulting cyclic peptide. Across six model variations, it was observed that increasing the \textit{pLDDT} weighting consistently improved generation results. Finally, leveraging insights from the previous explorations, a weighted \textit{pLDDT} metric was developed. When combined with a step-wise dynamic generation approach, this successfully generated cyclic peptides that mimic the reference molecule BMS-818251. This study demonstrates the potential for optimized precision generation of cyclic peptides targeting the HIV CD4-binding site and reveals methods for tuning computational models like AlphaFold to enhance the drug-discovery process for attachment inhibitors.

\clearpage
\section*{Acknowledgments}

This research project was conducted under the supervision of Prof. Nevin Zhang from the CSE department of Hong Kong University of Science and Technology and Dr. Tao Wang, inventor of HIV antiretroviral drug.

\clearpage
\appendix
\onecolumn
\section{Appendix 1: Hotspot Generation Time }
Given the long generation time of cyclic peptides interfacing with HIV’s gp120 trimer, one could also hypothesize that varying the interface contact scores could optimize the generation time. However, in an exploratory generation, varying hotspots had little impact on reducing the generation time. Hotspot regions of residues in increasing increments of 5 residues were configured during the generation run, and the generation time in seconds was measured.

\begin{figure}[H]
    \includegraphics[width=1\linewidth]{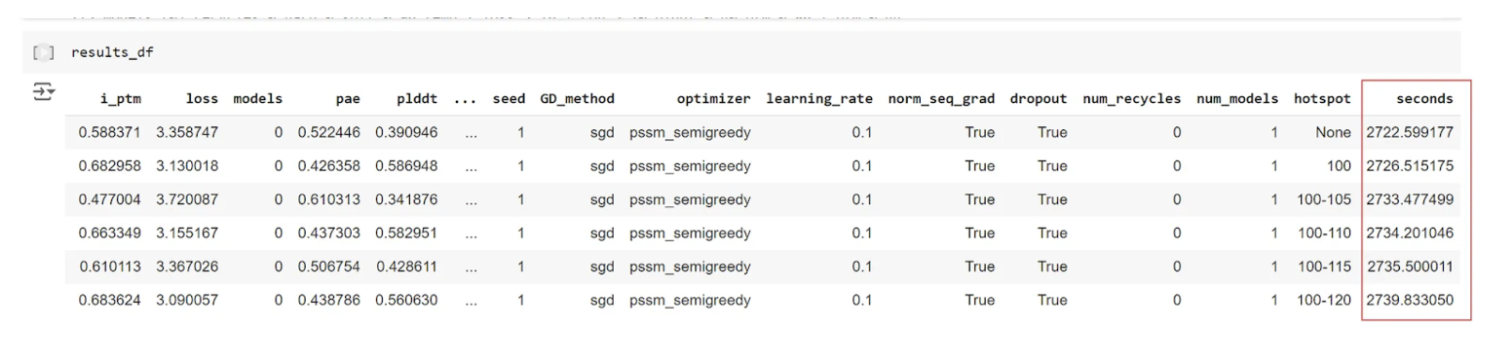}
    \caption{Generation time of varying hotspot configurations}
    \label{fig:app1f1}
\end{figure}

\clearpage
\section{Appendix 2: Results for each seed during the evaluation process of model loss functions}
\begin{figure}[H]
    \centering
    \includegraphics[width=0.75\linewidth]{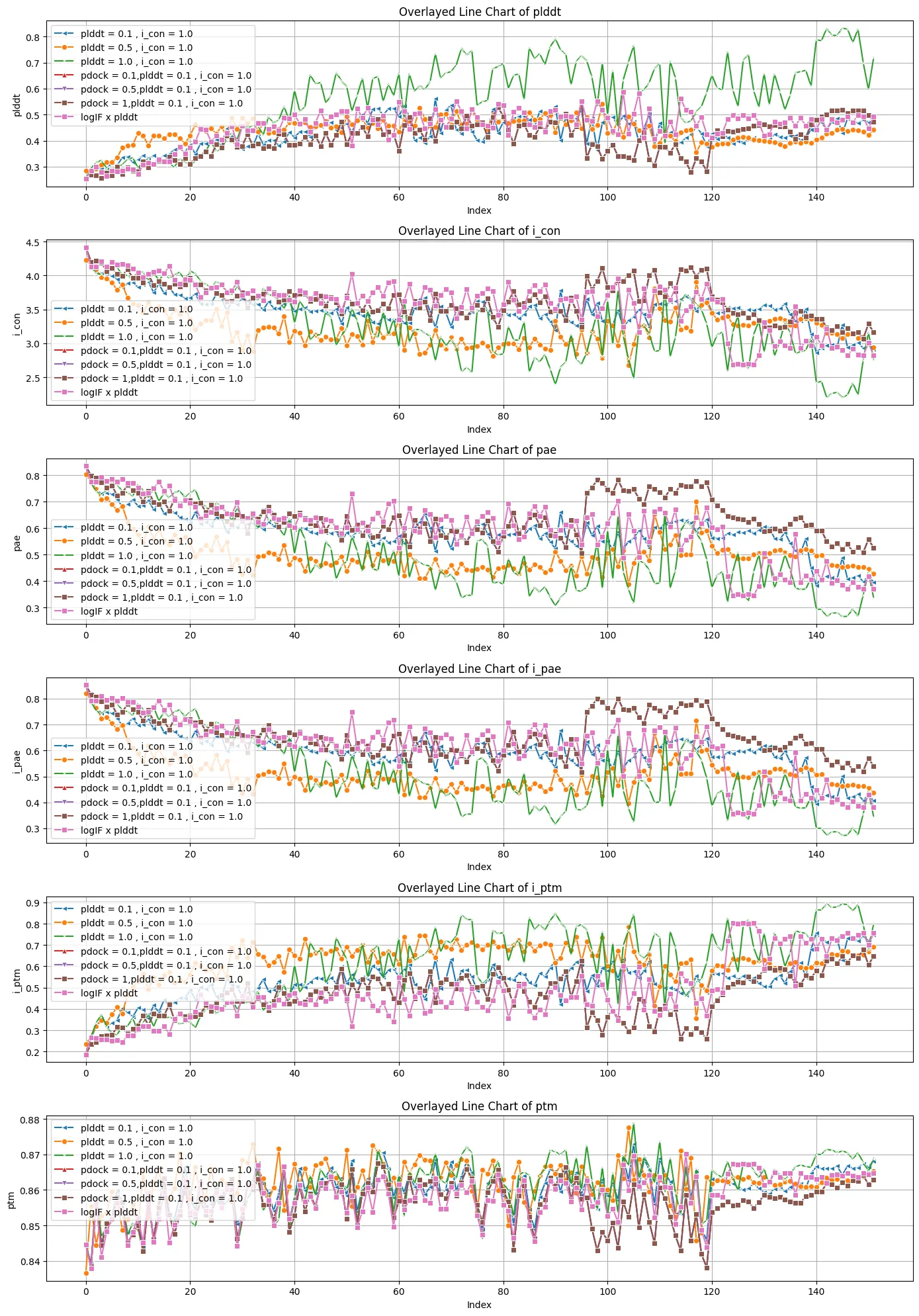}
    \caption{Seed20}
    \label{fig:A2.1}
\end{figure}

\begin{figure}[H]
    \centering
    \includegraphics[width=0.75\linewidth]{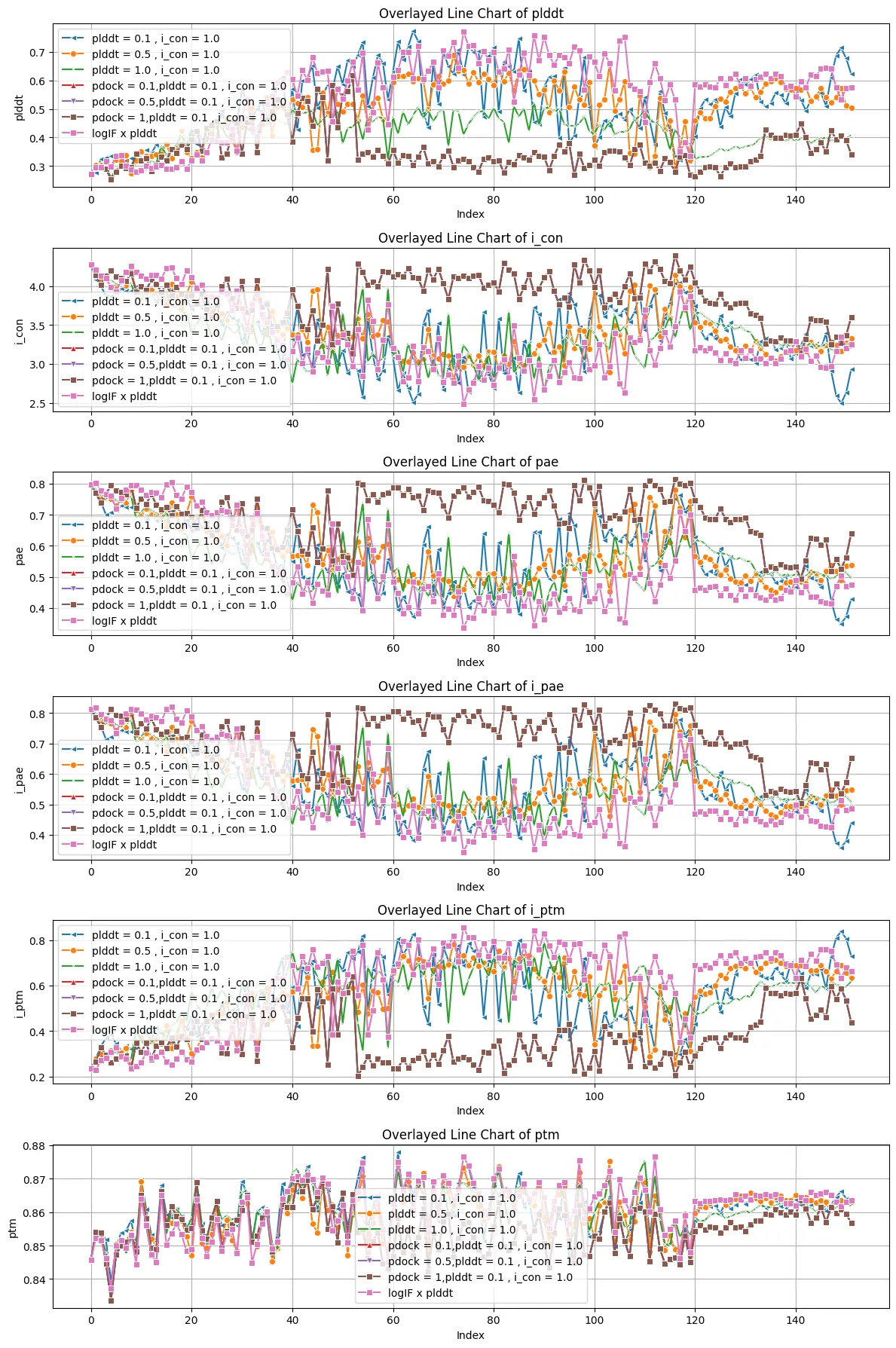}
    \caption{Seed23}
    \label{fig:A2.2}
\end{figure}

\begin{figure}[H]
    \centering
    \includegraphics[width=0.75\linewidth]{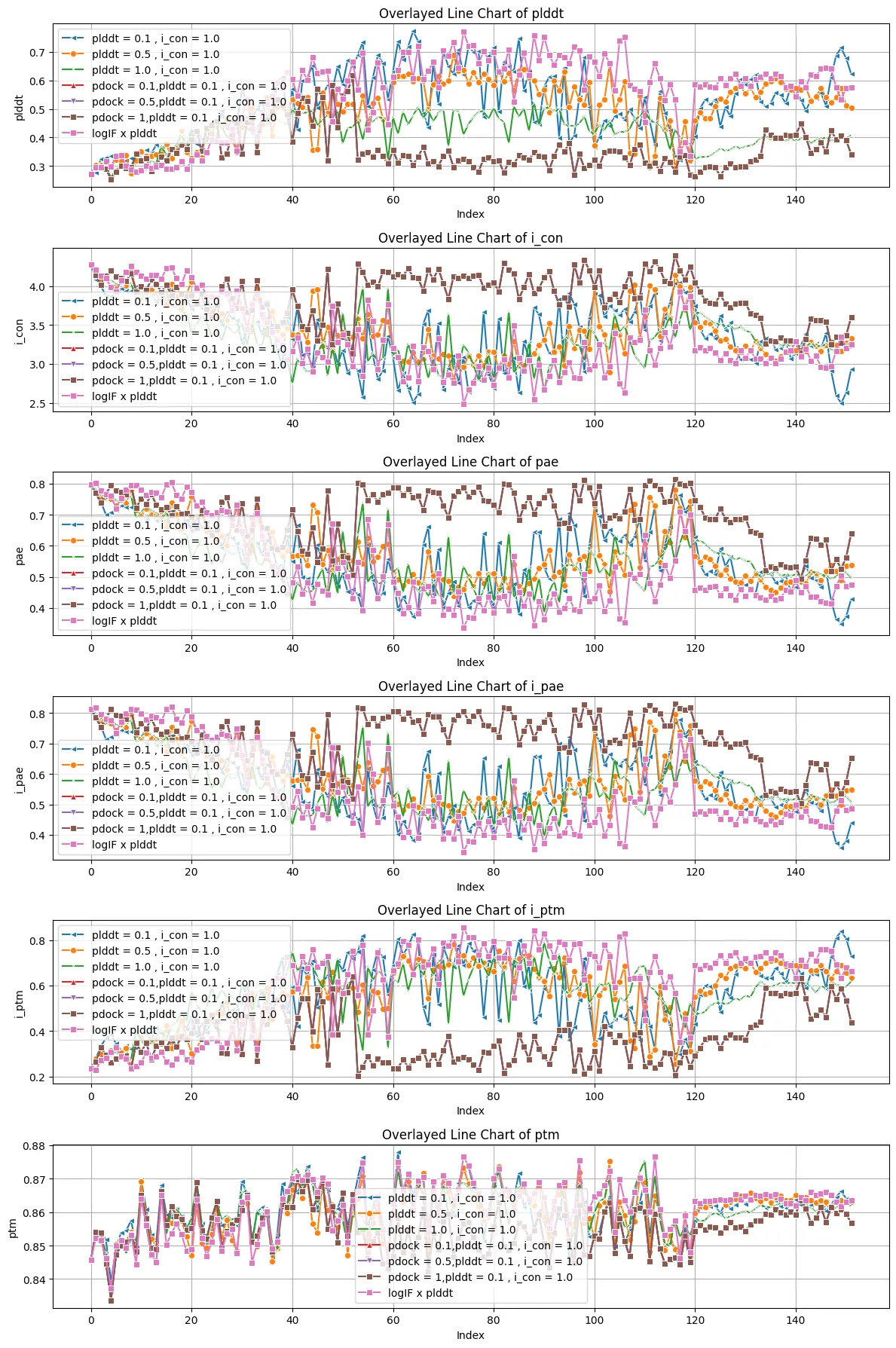}
    \caption{Seed25}
    \label{fig:A2.3}
\end{figure}

\begin{figure}[H]
    \centering
    \includegraphics[width=0.75\linewidth]{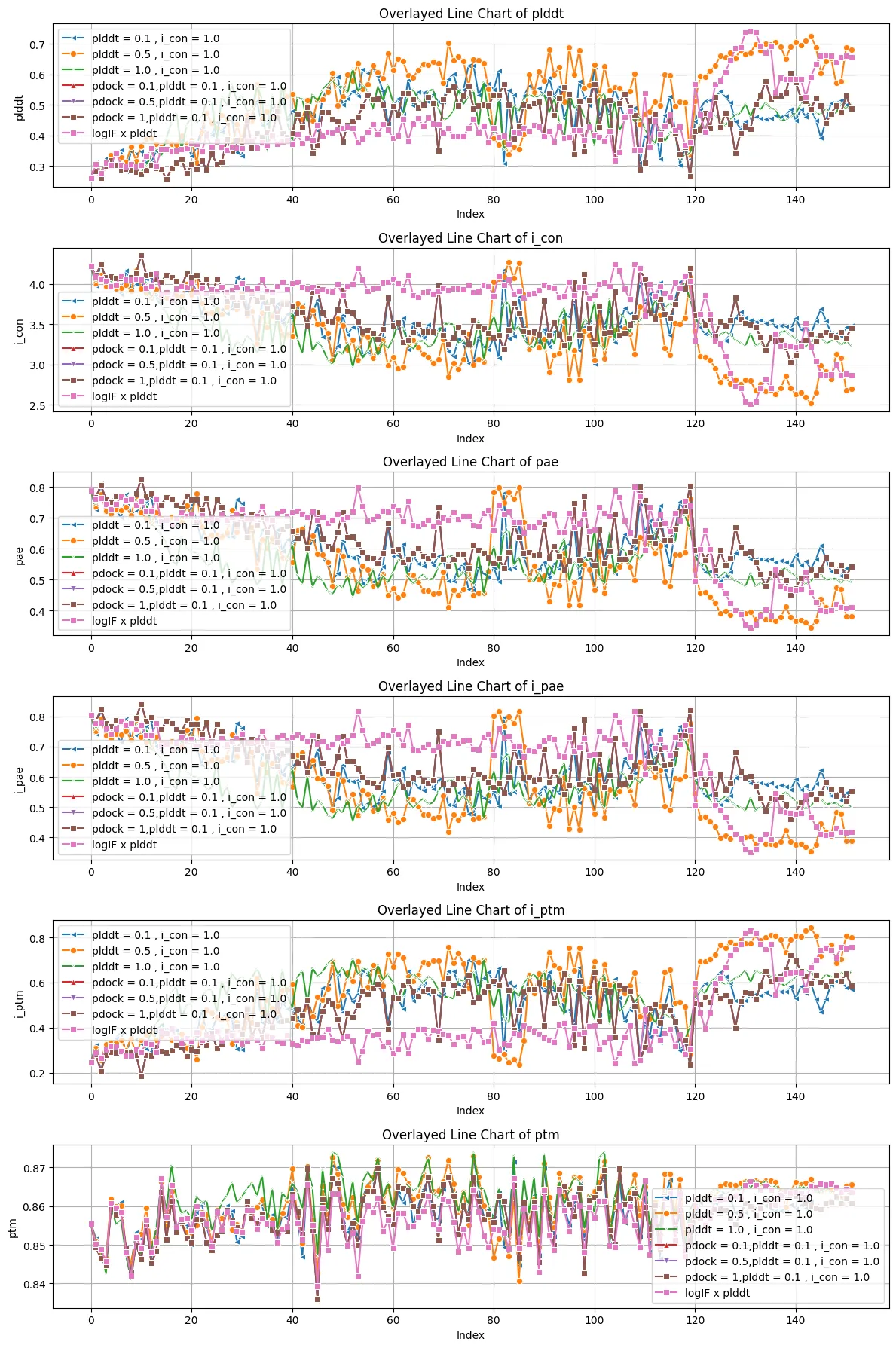}
    \caption{Seed26}
    \label{fig:A2.4}
\end{figure}
\begin{figure}[H]
    \centering
    \includegraphics[width=0.75\linewidth]{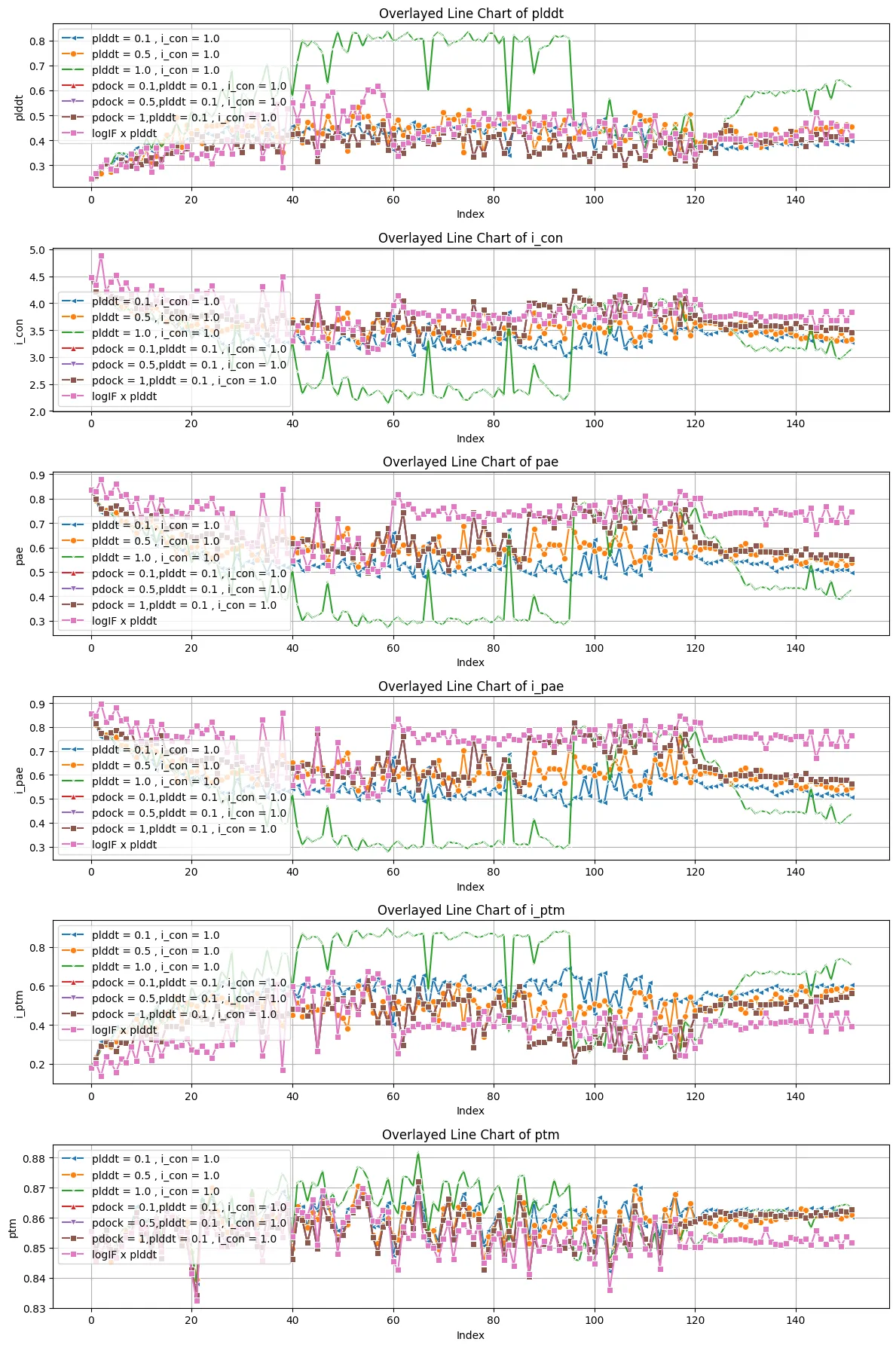}
    \caption{Seed27}
    \label{fig:A2.5}
\end{figure}

\clearpage
\section{Appendix 3: Normalized generation loss across 5 seeds}
\begin{figure}[H]
    \centering
    \includegraphics[width=0.75\linewidth]{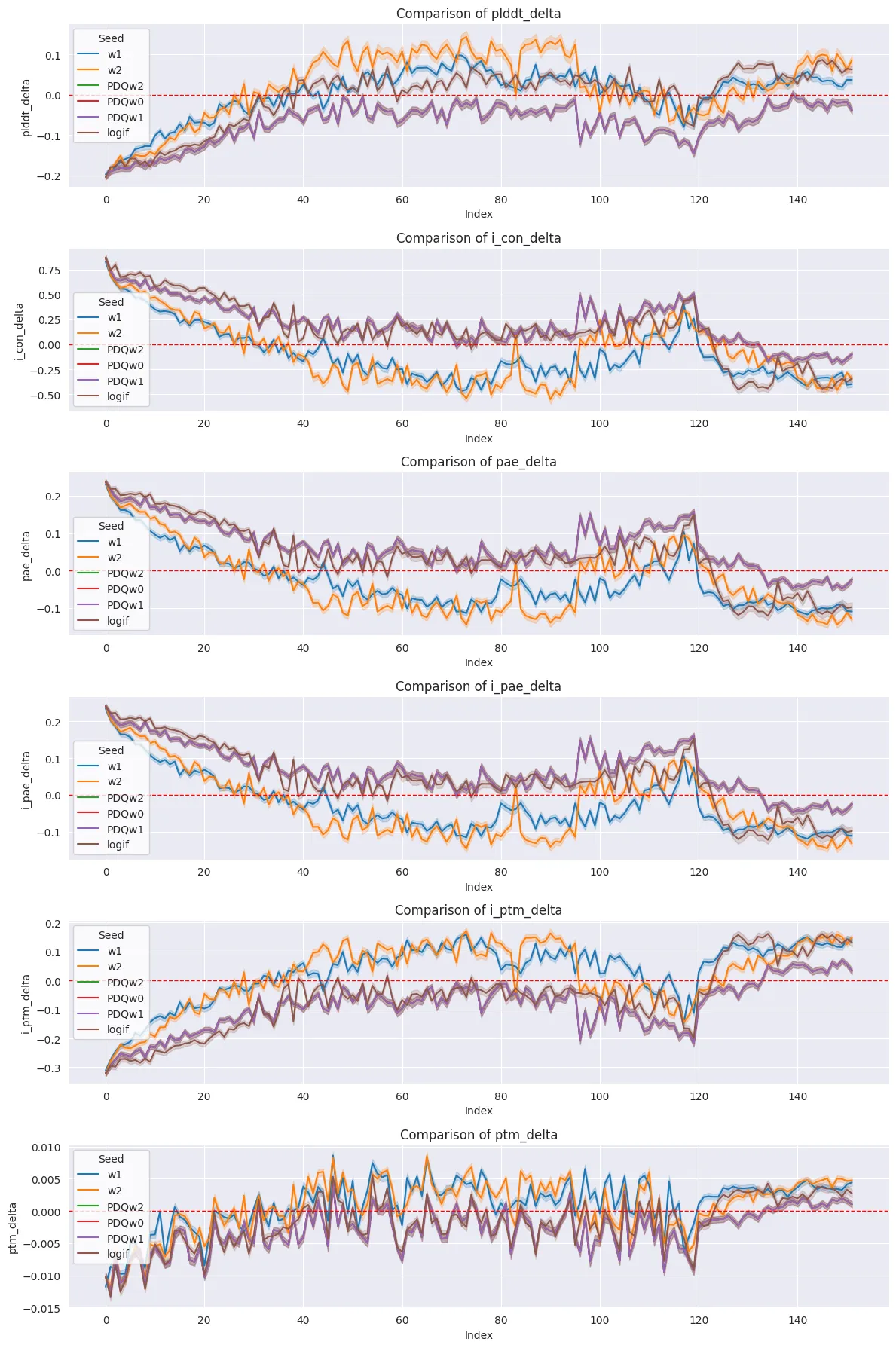}
    \caption{Normalized Results across 5 seeds}
    \label{fig:A3.1}
\end{figure}
\clearpage
\section{Appendix 4: Non-linear snapping of confidence during generation}
\begin{figure}[H]
    \centering
    \includegraphics[width=1\linewidth]{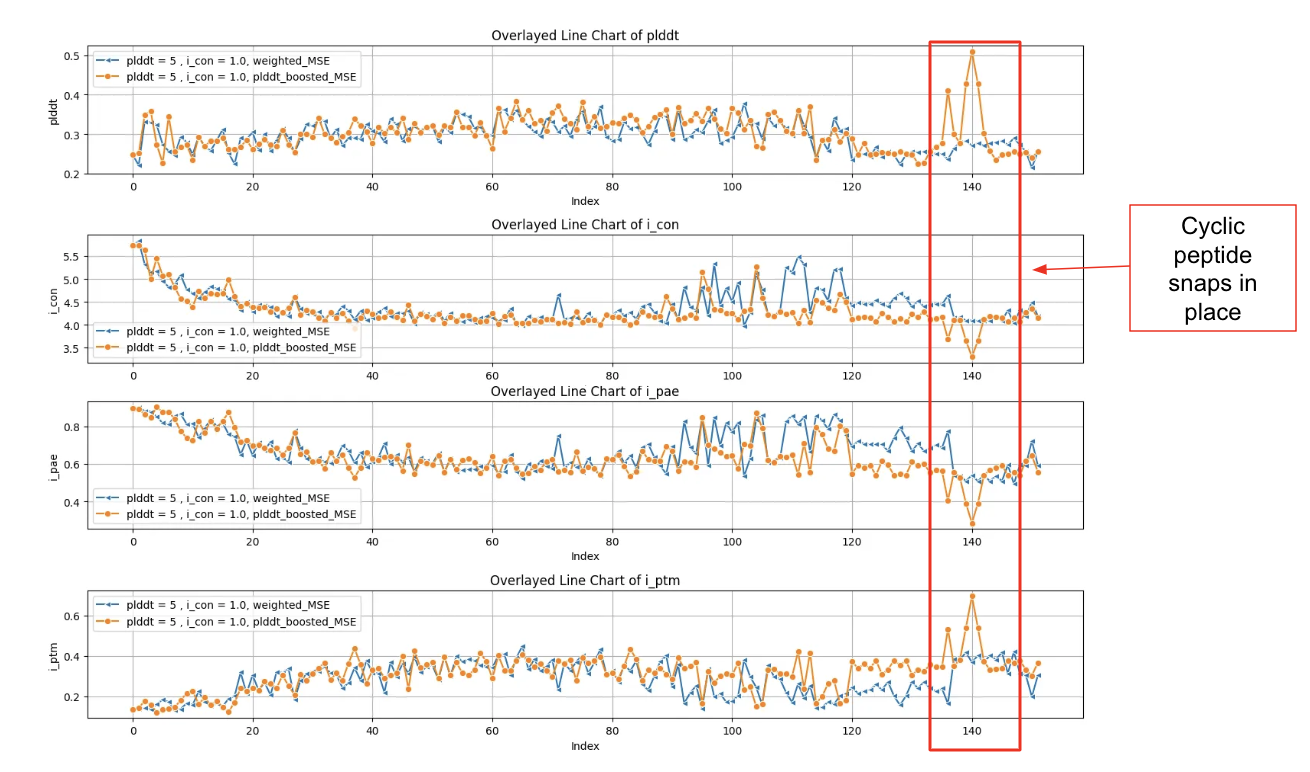}
    \caption{Normalized Results across 5 seeds}
    \label{fig:A3.2}
\end{figure}


\begin{thebibliography}{}

\bibitem[\protect\citename{Jumper et al.}2021]{Jumper:21}
Jumper, J., Evans, R., Pritzel, A., et al.
\newblock 2021.
\newblock Highly accurate protein structure prediction with AlphaFold.
\newblock {\em Nature}, 596, 583--589.
\newblock \url{https://doi.org/10.1038/s41586-021-03819-2}.

\bibitem[\protect\citename{Pancera et al.}2017]{Pancera:17}
Pancera, M., Lai, Y. T., Bylund, T., et al.
\newblock 2017.
\newblock Crystal structures of trimeric HIV envelope with entry inhibitors BMS-378806 and BMS-626529.
\newblock {\em Nature Chemical Biology}, 13(11), 1115--1122.
\newblock \url{https://doi.org/10.1038/nchembio.2460}.

\bibitem[\protect\citename{Lynch et al.}2012]{Lynch:12}
Lynch, R. M., Tran, L., Louder, M. K., Schmidt, S. D., Cohen, M., CHAVI 001 Clinical Team Members, Dersimonian, R., Euler, Z., Gray, E. S., Abdool Karim, S., Kirchherr, J., Montefiori, D. C., Sibeko, S., Soderberg, K., Tomaras, G., Yang, Z. Y., Nabel, G. J., Schuitemaker, H., Morris, L., Haynes, B. F., \& Mascola, J. R.
\newblock 2012.
\newblock The development of CD4 binding site antibodies during HIV-1 infection.
\newblock {\em Journal of Virology}, 86(14), 7588--7595.
\newblock \doi{10.1128/JVI.00734-12}.
\newblock Epub 2012 May 9.
\newblock PMID: 22573869; PMCID: PMC3416294.

\bibitem[\protect\citename{Abramson et al.}2024]{Abramson:24}
Abramson, J., Adler, J., Dunger, J., et al.
\newblock 2024.
\newblock Accurate structure prediction of biomolecular interactions with AlphaFold 3.
\newblock {\em Nature}, 630, 493--500.
\newblock \url{https://doi.org/10.1038/s41586-024-07487-w}.

\bibitem[\protect\citename{Rettie et al.}2023]{Rettie:23}
Rettie, S. A., Campbell, K. V., Bera, A. K., Kang, A., Kozlov, S., De La Cruz, J., Adebomi, V., Zhou, G., DiMaio, F., Ovchinnikov, S., Bhardwaj, G.
\newblock 2023.
\newblock Cyclic peptide structure prediction and design using AlphaFold.
\newblock {\em bioRxiv [Preprint]}.
\newblock doi: 10.1101/2023.02.25.529956.

\bibitem[\protect\citename{Kosugi and Ohue}2023]{Kosugi:23}
Kosugi, T. and Ohue, M.
\newblock 2023.
\newblock Design of cyclic peptides targeting protein-protein interactions using AlphaFold.
\newblock {\em Int J Mol Sci}, 24(17):13257.
\newblock doi: 10.3390/ijms241713257.

\bibitem[\protect\citename{Bryant et al.}2022]{Bryant:22}
Bryant, P., Pozzati, G., and Elofsson, A.
\newblock 2022.
\newblock Improved prediction of protein-protein interactions using AlphaFold2.
\newblock {\em Nat Commun}, 13, 1265.
\newblock \url{https://doi.org/10.1038/s41467-022-28865-w}.

\bibitem[\protect\citename{Mirdita et al.}2022]{Mirdita:22}
Mirdita, M., Schütze, K., Moriwaki, Y., et al.
\newblock 2022.
\newblock ColabFold: making protein folding accessible to all.
\newblock {\em Nat Methods}, 19, 679--682.
\newblock \url{https://doi.org/10.1038/s41592-022-01488-1}.

\bibitem[\protect\citename{Lai et al.}2019]{Lai:19}
Lai, Y. T., Wang, T., O'Dell, S., Louder, M. K., Schön, A., Cheung, C. S. F., Chuang, G. Y., Druz, A., Lin, B., McKee, K., Peng, D., Yang, Y., Zhang, B., Herschhorn, A., Sodroski, J., Bailer, R. T., Doria-Rose, N. A., Mascola, J. R., Langley, D. R., Kwong, P. D.
\newblock 2019.
\newblock Lattice engineering enables definition of molecular features allowing for potent small-molecule inhibition of HIV-1 entry.
\newblock {\em Nat Commun}, 10(1):47.
\newblock doi: 10.1038/s41467-018-07851-1.

\bibitem[\protect\citename{Grant and Kozal}2022]{Grant:22}
Grant, P. M. and Kozal, M. J.
\newblock 2022.
\newblock Fostemsavir: a first-in-class HIV-1 attachment inhibitor.
\newblock {\em Curr Opin HIV AIDS}, 17(1):32--35.
\newblock doi: 10.1097/COH.0000000000000712.

\bibitem[\protect\citename{Alford et al.}2017]{Alford:17}
Alford, R. F., Leaver-Fay, A., Jeliazkov, J. R., O'Meara,
\end{thebibliography}
\end{document}